\documentclass{emulateapj}
\usepackage{psfig}

\newcommand{\be}{\begin{equation}}
\newcommand{\ee}{\end{equation}}

\slugcomment{Submitted for publication to ApJ}

\shorttitle{Wavelet morphology}
\shortauthors{Mart{\'\i}nez et al.}

\begin{document}

\title{Morphology of the galaxy distribution from wavelet denoising}
\author{Vicent J. Mart{\'\i}nez}
\affil{Observatori Astron\`omic, Universitat de Val\`encia, Apartat
de Correus 22085, E-46071 Val\`encia, Spain} \email{martinez@uv.es}

\author{Jean-Luc Starck}
\affil{CEA-Saclay, DAPNIA/SEDI-SAP, Service d'Astrophysique, F-91191 Gif
sur Yvette, France}
\email{jstarck@cea.fr}

\author{Enn Saar}
\affil{Tartu Observatoorium, T\~oravere, 61602, Estonia}
\email{saar@aai.ee}

\author{David L. Donoho}
\affil{Department of Statistics, Stanford University, Sequoia Hall,
Stanford, CA 94305, USA}
\email{donoho@stat.stanford.edu}

\author{Simon Reynolds}
\affil{Blackett Laboratory, Imperial College London, SW7 2AZ, UK}
\email{simon.reynolds@imperial.ac.uk}

\author{Pablo de la Cruz}
\affil{Observatorio Astron\'omico, Universidad de Valencia, Apartado
de Correos 22085, E-46071 Valencia, Spain} 
\email{pablo.cruz@uv.es}

\and

\author{Silvestre Paredes}
\affil{Departamento de Matem\'atica Aplicada y Estad\'{\i}stica,
Universidad Polit\'ecnica de Cartagena, 30203 Cartagena, Spain }
\email{silvestre.paredes@upct.es}

\begin{abstract}
We have developed a method based on wavelets to obtain the true
underlying smooth density from a point distribution. The goal has
been to reconstruct the density field in an optimal way ensuring
that the morphology of the reconstructed field reflects the true
underlying morphology of the point field which, as the galaxy
distribution, has a genuinely multiscale structure, with
near-singular behavior on sheets, filaments and hotspots.
If the discrete distributions are smoothed using Gaussian
filters, the morphological properties tend to be closer to
those expected for a Gaussian field. The use of wavelet denoising
provide us with a unique and more accurate morphological description.
\end{abstract}


\section{Introduction}

The large-scale structure of the universe shows intricate patterns
with filaments, clusters, and sheet-like arrangements of galaxies
encompassing large nearly empty regions, the so-called voids. This
complex structure shows clearly non-Gaussian features. However, it
is likely that the observed structure developed from tiny
fluctuations of an initial Gaussian random field by the action of
gravity. This is the scenario suggested by the analysis of the maps
of the microwave background radiation. Thus, it is important to
check if the present large-scale structure is compatible with the
Gaussianity of the initial fluctuations.

Different statistical measures have been used in the cosmological
literature to quantitatively describe the cosmic texture
\citep{cf:martinezsaar}. To complement the information provided by
the second-order descriptors --the correlation function and the
power spectrum-- different alternatives have been proposed. Some of
these statistics are focused in quantifying geometrical and
morphological aspects of the distribution. In this context, the {\it
genus}, introduced to measure deviations from Gaussianity
\citep{gott86}, is one of the most widely used techniques. The genus
and its generalization, the Minkowski functionals, allow us to
quantify the morphology of the isodensity surfaces of the matter
distribution.

The Minkowski functionals describe the morphology of hypersurfaces,
with dimensionality one less than that of the encompassing space. In
the analysis of the three-dimensional matter distribution, the
functionals are applied to isodensity surfaces separating regions
with density above and below a given threshold. This implies that
the first step is to obtain a smooth density field from the discrete
distribution of matter. The morphological descriptors can be applied
both to the observed galaxy distribution and to the dark-matter
based $N$-body simulations of the large-scale structure. In all
cases, we have to smooth the data to construct a real density field.
This smoothing has to be more severe, when we want to measure the
morphology of a discrete point distribution as a redshift catalog,
than when we measure the morphology of the dark matter distribution
in cosmological simulations.

The first assumption we have to make is that the galaxy distribution
is a sample from a Cox process (see, e.g., \citet{cf:martinezsaar});
the galaxy positions $(x_i,y_i,z_i), i=1,...,n$ represent a point
process which samples the continuous field. Smoothing has to
reconstruct the underlying $f(x,y,z)$, and if the smoothing is done
either well or poorly, then the estimated field $\hat{f}$ will be
either good or bad at representing the true underlying field $f$.

It is well known that when we are estimating a density field $f$,
there is a critical smoothing level at which it begins to be true
that the estimated field resembles the true field.  For example,
with a $C^2$ density in dimension 1, we need to smooth with the
bandwidth
\begin{equation}
h_n \sim a n^{-1/5},
\label{band}
\end{equation}
where $a$ depends on the underlying density field $f$
\citep{donoho88}. If we smooth less than this,  $h < h_n$, then the
number of modes of the estimate $\hat{f}$ will tend to infinity,
while if we smooth more than this, the estimate will have fewer
modes than the true density \citep{silverman81}.

Cosmological density fields, instead of being a generic $C^2$ field,
are more complex in nature. They have a genuinely multiscale
structure, with near-singular behavior in sheets, filaments and
clusters. The smoothing procedures that are proper for such objects
are presumably entirely different than the smoothing that is good
for $C^2$ objects, and so we do not expect that Eq. \ref{band} can
be applied in this setting.

Correct smoothing should also be spatially adaptive, so that locally
it is using a scale based on the degree of smoothness of the object,
or the scale should be smaller than the statistically significant
structures. The method advocated in this paper, based on wavelet
thresholding, does this automatically and provides a smoothing
recipe that is unique for a given realization of a point process and
does not depend on an a priori chosen bandwidth.

Having obtained a consistent estimate of the density field, we can
be certain that the morphology of the reconstructed field reflects
the true underlying morphology of the point field. The goal of the
present paper is finding and analyzing such morphological
descriptors.

\section{Smoothing schemes}

In this section we will introduce two different smoothing techniques
that can be applied to obtain a continuous density field from a
discrete point distribution. Our goal is to analyze how these
schemes affect the correct determination of the Minkowski
functionals, and which is the best to study the morphology of the
matter distribution.

\subsection{Gaussian smoothing}

For morphological studies, smoothing is typically done by using a
Gaussian kernel \be W({\bf x})= \frac{1}{(2\pi)^{3/2} \sigma^3} \exp
\left ( - \frac{{\bf x}^2}{2\sigma^2} \right ). \ee

The window width $\sigma$ is the parameter that governs the level of
smoothing of the discrete data to obtain the kernel density
estimate. \citet{ham86} recommend that the smoothing length has to
be chosen larger than the correlation length, $r_0$, the distance at
which the two-point correlation function $\xi(r_0)=1$. This is the
recipe that is usually used for the morphological analysis of the
observed galaxy distribution, together with a requirement that the
smoothing length should also be larger than the typical size of the
volume-per-galaxy\footnote{$d$ is typically referred to as the mean
interparticle separation.}  $d=(V/n)^{1/3}$, where $V$ is the total volume of
the sample and $n$ the number of points (galaxies) (see, e.g.,
\citet{hoyle02}).

A lot of work has been done by the statistical community on the
optimal smoothing length that would give the best density estimate.
However, as \citet{silverman81} has pointed out: ``Most methods seem
to depend on some arbitrary choice of the scale of the effects being
studied". Certainly to choose the appropriate value of $\sigma$ is
an art, but in any case we must avoid two kinds of artifacts:
undersmoothing, which causes huge numbers of spurious oscillations
and oversmoothing, which removes real features of structure. This
last aspect is crucial when measuring the morphology of the large
scale structure because, since smoothing has to be large enough to
describe morphology reliably, it will inevitably erase small-scale
non-Gaussian features. \citet{coles95} note that ``smoothing on
scales much larger than the scale at which correlations are
significant will tend to produce a Gaussian distribution by virtue
of the central limit theorem" \citep{martinez93}. A conservative
approach is based on searching for efficient and consistent estimates of
the bandwidth that are typically upper bounds. These scales would
reveal as much detail as the optimal bandwidth, if it exists
\citep{donoho88}.

\subsection{Wavelet denosing}
The Undecimated Isotropic Wavelet Transform (UIWT), also 
named {\em \`a trous algorithm}, 
decomposes an $n \times n \times n$ data set $D$ as 
a superposition of
the form
\[
D = {c_J} + \sum_{j=1}^{J} w_{j},
\]
where $c_{J}$ is a coarse or smooth version of the original data $D$
and $w_j$ represents the details of $D$ at scale $2^{-j}$ (see 
\citet{starck:book98,starck:book02} for details).  Thus, the
algorithm outputs $J+1$ sub-band arrays of size $n \times n \times n$. We will
use an indexing convention such that $j = 1$ corresponds to the finest
scale (high frequencies). 
Wavelets have been used successfully  for   
denoising  via non-linear filtering or 
thresholding methods \citep{starck:book02}. 
Hard thresholding, for instance, consists of setting 
all insignificant coefficients (\emph{i.e.} 
coefficients with an absolute value below a given threshold) to zero.
  
For the noise model, given that this relates to point pattern clustering, 
we have to consider the Poisson noise case. The autoconvolution histogram 
method \citep{slezak} used for X-ray image \citep{stapie,pierre04,ivan04}
can also be used here.
It consists of calculating numerically the probability distribution function (pdf) of a wavelet $w_{j,x,y,z}$
coefficient with the hypothesis that the galaxies used for obtaining $w_{j,x,y,z}$ are 
randomly distributed. The pdf is obtained by autoconvolving $n$ times the histogram of the wavelet function,
$n$ being the number of galaxies which have been used for obtaining $w_{j,x,y,z}$, i.e. the number 
of galaxies in a box around $(x,y,x)$, the size of the box depending on the scale $j$.
More details can be found in \citet{stapie,starck:book02}.

Once the pdf relative to the coefficient $w_{j,x,y,z}$ is known, we can detect the significant 
wavelet coefficients easily. We derive two threshold values 
$T^{min}_{j,x,y,z}$ and $T^{max}_{j, x,y,z}$ such that 
\begin{eqnarray}
\mbox{Prob} (W < T^{min}_{j,x,y,z}) & = & \epsilon \nonumber \\
\mbox{Prob} (W > T^{max}_{j,x,y,z}) & = & \epsilon
\label{TestHyp}
\end{eqnarray}
$\epsilon$ corresponding to the confidence level, 
and the positive (respective negative) wavelet coefficent is significant if it is larger 
than $T^{max}_{j,x,y,z}$ (resp. lower than $T^{min}_{j,x,y,z}$).
Denoting $D$ the noisy data and $\delta$ the thresholding operator, the filtered data $\tilde D$ are obtained by : 
\begin{eqnarray}
 {\tilde D} =    {\cal R} \delta( {\cal T} D)
\end{eqnarray}
where ${\cal T}$ is the wavelet transform operator and ${\cal R}$ is 
the wavelet reconstruction operator. 
In practice, we get better results using the iterative reconstruction described in \citet{starck:book02}
which minimizes the $l_1$ norm of the wavelet coefficients. It is this iterative technique 
that we have used for our experiments.
 
Poisson noise denoising has been addressed in a series of recent papers
\citep{fryz1,kola1,kola4,nowa2,anto1,timm1,jammal04}. All of them uses the
Haar wavelet transform because it presents the interesting property that
a Haar wavelet coefficient is the difference between two variables which follow
a Poisson distribution. This property allows us to derive an analytical form of 
the pdf of the wavelet coefficients. The Haar transform has however several 
drawbacks such block artifact creation or a tendency to create square structures.
For XMM, it was also shown than the  isotropic wavelet transform 
was much more powerful for  detecting clusters of galaxies  \citep{ivan01}. 

\section{Morphological descriptors}

\subsection{The genus curve}

Historically, the first morphological descriptor used was the genus
\citep{gott86}. The genus $G(S)$ measures the connectivity of a
surface, $S$, with holes and disconnected pieces, by the difference
of the number of holes and the number of isolated regions:
\[
G(S) = \mbox{number of holes} - \mbox{number of isolated
regions}+1.
\]
The genus of a sphere is $G=0$, a torus or a sphere with a handle
have the genus $G=+1$, a sphere with $N$ handles has the genus
$G=+N$, while the collection of $N$ disjoint spheres has the genus
$G=-(N-1)$. The genus describes the topology of the isodensity
surfaces, thus its study is in the cosmological literature
frequently called ``topological analysis''.

The genus curve is usually parameterized by two related quantities,
the filling factor, $f$, which is the fraction of the survey volume
above the density threshold or, alternatively, by the quantity $\nu$
defined by \be \label{eq:nu} f = \frac{1}{\sqrt{2\pi}}
\int_\nu^\infty e^{-t^2/2}dt. \ee In the case of a Gaussian random
field, $\nu$ is also the number of standard deviations by which the
threshold density departs from the mean density, and with this
parametrization, the genus per unit volume of a surface, $S$,
corresponding  to a given density threshold, $g\equiv (G(S)-1)/V$,
follows the analytical expression \be g(\nu) = N (1-\nu^2) \exp
\left ( - \frac{\nu^2}{2} \right ), \ee where the amplitude $N$
depends on the power spectrum of the random field \citep{ham86}. If
the density distribution is not Gaussian, the parameterization
(\ref{eq:nu}) eliminates the (trivial) non-Gaussianity caused by the
one-point density distribution. Opinions differ about which argument
is better; we shall use $\nu$ in this paper.

\begin{figure*}
\centering
\resizebox{\textwidth}{!}{\includegraphics*{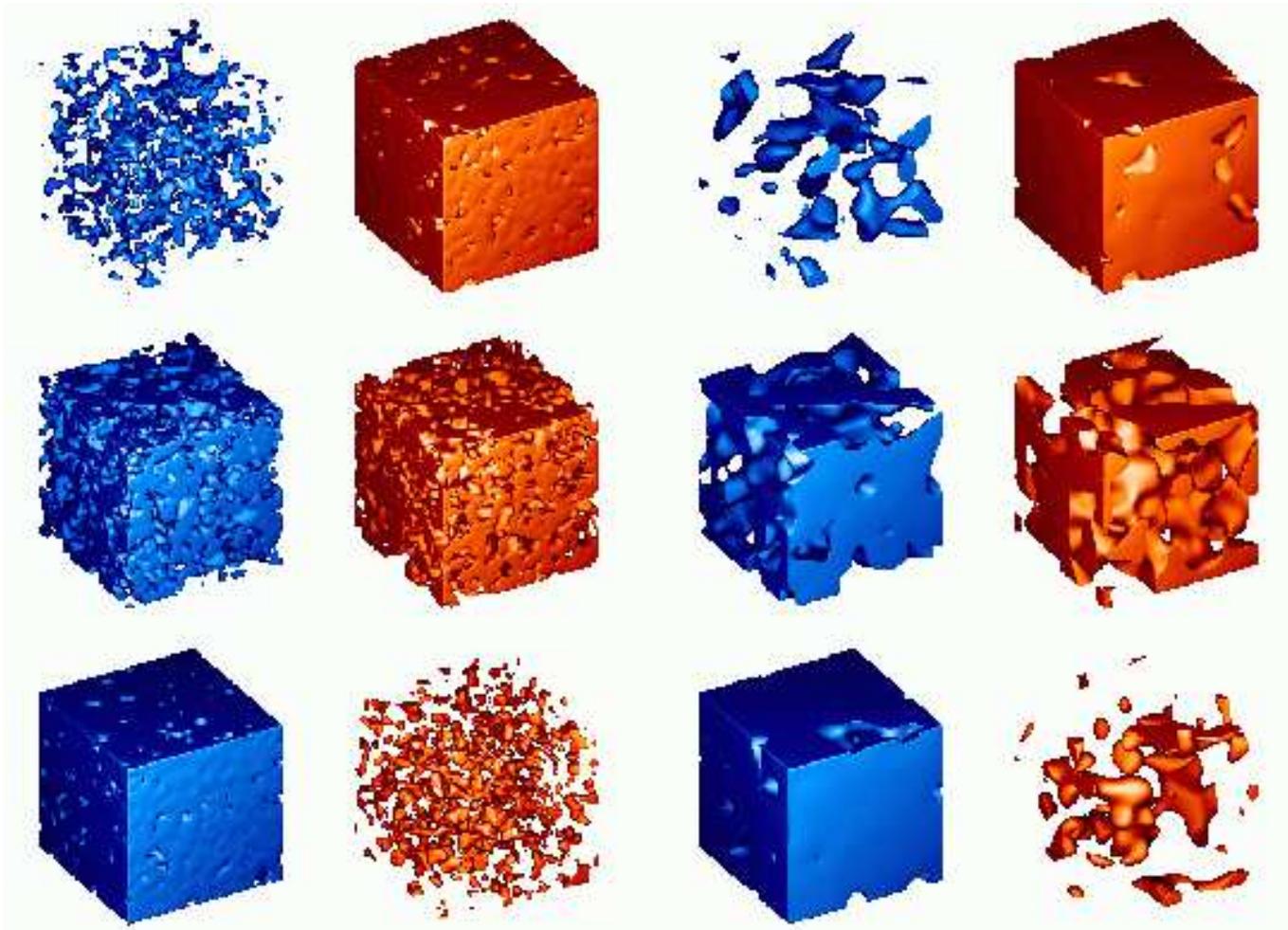}}
\caption{The two columns on the left show the spatial distribution
of the low- (first column) and high density (second column) regions
for a realization of a Gaussian random field, with comparatively
little smoothing $(\sigma=3$). The upper pair shows the 7\% (volume
fraction) low, 93\% high density regions, the middle pair stands for
50\%--50\%, and the lower pair shows the 93\% low-density, 7\%
high-density case. The two columns on the right are the same, but
for heavy smoothing $(\sigma=8$). \label{fig:gaussden}}
\end{figure*}

This curve, symmetric about 0 in $\nu$, is typical of the
random-phase morphology. We have simulated a Gaussian random field
with a power-law spectrum $P(k)\sim k^{-1}$ and this field has been
smoothed with a Gaussian kernel with $\sigma=3$ (the cube size is
128). As we see in Fig.~\ref{fig:gaussden} (left panels), the
regions with density above or below the mean value are statistically
indistinguishable. In the right column of this figure we show the
isodensity surfaces for our realization, which encompass the denser
regions of the simulated box. The three panels, from top to bottom,
correspond respectively to 7\%, 50\%, and 93\% of the volume
encompassing regions with higher density. Likewise, the left column
shows the low-density regions corresponding to the same percentage
of the volume. The symmetry between the high-density and the
low-density regions is clearly seen. The right panels of
Fig.~\ref{fig:gaussden} depict the same realization, but more
heavily smoothed, with the smoothing length $\sigma=8$. These are
the standard distributions, which are typically compared with
observational data. Such a morphology is usually called ``the sponge
morphology''.  The sponginess of the isodensity surfaces is clearly
seen, particularly at the central pair of panels, in both figures,
corresponding to the 50\% low and high densities: the surface
separating both regions has many holes, is multiply connected, and
has negative curvature.

Other types of genus curves can be found in the cosmological
literature. When rich clusters dominate the distribution, the genus
curves are shifted to the left, and the morphology is referred to as
``meat-ball'', while the expression ``Swiss-cheese'' is used for
right-shifted genus curves corresponding to distributions with empty
bubbles surrounded by a single high density region.

\subsection{Minkowski functionals}

An elegant generalization of the genus statistic is to consider this
measure as one of the four Minkowski functionals which describe
different morphological aspects of the galaxy distribution
\citep{cf:mecke94}. These functionals provide a complete family of
morphological measures -- all additive, motion invariant and
conditionally continuous functionals defined for any hypersurface
are  linear combinations of its Minkowski functionals.

The Minkowski functionals (MF for short) describe the morphology of
isodensity surfaces \citep{minkowski,tomita}, and depend thus on two factors -- the smoothing
procedure and the specific density level, (see \citet{sheth05} for a
recent review). An alternative approach starts from the point field,
decorating the points with spheres of the same radius, and studying
the morphology of the resulting surface \citep{ker96var,ker2}. These
functionals depend only on one parameter (the radius of the
spheres), but this approach does not refer to a density; we shall
not use that for the present study.

The Minkowski functionals are defined as follows. Consider an
excursion set $F_\phi$ of a field $\phi(\mathbf{x})$ in 3-D (the set
of all points where $\phi(\mathbf{x}\ge\phi$). Then, the first
Minkowski functional (the volume functional) is the volume of the
excursion set:
\[
V_0(\phi)=\int_{F_\phi}d^3x.
\]
The second MF is proportional to the surface area
of the boundary $\delta F_\phi$ of the excursion set:
\[
V_1(\phi)=\frac16\int_{\delta F_\phi}dS(\mathbf{x}).
\]
The third MF is proportional to the integrated mean curvature
of the boundary:
\[
V_2(\phi)=\frac1{6\pi}\int_{\delta F_\phi}
    \left(\frac1{R_1(\mathbf{x})}+\frac1{R_2(\mathbf{x})}\right)dS(\mathbf{x}),
\]
where $R_1$ and $R_2$ are the principal curvatures of the boundary.
The fourth Minkowski functional is proportional to the integrated
Gaussian curvature (the Euler characteristic) of the boundary:
\[
V_3(\phi)=\frac1{4\pi}\int_{\delta F_\phi}
    \frac1{R_1(\mathbf{x})R_2(\mathbf{x})}dS(\mathbf{x}).
\]
The last MF is simply related to the morphological genus $g$
introduced in the previous subsection by
\[
V_3=\chi=\frac12(1-G)
\]
($\chi$ is the usual notation for the Euler characteristic). The
functional $V_3$ is a bit more comfortable to use -- it is additive,
while $G$ is not, and it gives just twice the number of isolated
balls (or holes). Although the genus continues to be widely used, in
several recent papers many authors have chosen to present the
Minkowski functional $V_3$; we shall follow this recent and logical
trend. Instead of the functionals, their spatial densities $V_i$ are
frequently used:
\[
v_i(f)=V_i(f)/V, \quad i=0,\dots,3,
\]
where $V$ is the total sample volume.

All the Minkowski functionals have analytic expressions for
isodensity slices of realizations of Gaussian random fields.
For three-dimensional space they are:
\begin{eqnarray*}
v_0&=&\frac12-\frac12\Phi\left(\frac{\nu}{\sqrt2}\right),\\
v_1&=&\frac23\frac{\lambda}{\sqrt{2\pi}}\exp\left(-\frac{\nu}2\right),\\
v_2&=&\frac23\frac{\lambda^2}{\sqrt{2\pi}}\nu\exp\left(-\frac{\nu}2\right),\\
v_3&=&\frac{\lambda^3}{\sqrt{2\pi}}(\nu^2-1)\exp\left(-\frac{\nu}2\right),
\end{eqnarray*}
where $\Phi(\cdot)$ is the Gaussian error integral, and $\lambda$
is determined by the correlation function $\xi(r)$ of the field as:
\[
\lambda^2=\frac1{2\pi}\frac{\displaystyle\xi''(0)}{\displaystyle\xi(0)}.
\]

\subsection{Numerical algorithms}

Several algorithms are used to calculate the Minkowski functionals
for a given density field and a given density threshold.  We can
either try to follow exactly the geometry of the isodensity surface,
e.g., using triangulation \citep{surfgen}, or to approximate the
excursion set on a simple cubic lattice. The algorithm that was
proposed first by \citet{gott86}, uses a decomposition of the field
into filled and empty cells, and another popular algorithm
\citep{coles96} uses a grid-valued density distribution. The
lattice-based algorithms are simpler and faster, but not as accurate
as the triangulation codes. The main difference is in the edge
effects -- while surface triangulation algorithms do not suffer from
these, edge effects may be rather serious for the lattice
algorithms.

        We use a simple grid-based algorithm, based
on integral geometry (the Crofton's intersection formula, see
\citet{jens97}). We find the density thresholds for given filling
fractions by sorting the grid densities, first. Vertices with higher
densities than the threshold form the excursion set. This set is
characterized by its basic sets of different dimensions -- points
(vertices), edges formed by two neighboring points, squares (faces)
formed by four edges, and cubes formed by six faces. The algorithm
counts the numbers of all basic sets, and finds the values of the
Minkowski functionals as
\begin{eqnarray*}
V_0(f)&=&a^3N_3,\\
V_1(f)&=&a^2\left(\frac29N_2(f)-\frac23N_3(f)\right),\\
V_2(f)&=&a\left(\frac29N_1(f)-\frac49N_2(f)+\frac23N_3(f)\right),\\
V_3(f)&=&N_0(f)-N_1(f)+N_2(f)-N_3(f),
\end{eqnarray*}
where $a$ is the grid step, $f$ is the filling factor, $N_0$ is the
number of vertices, $N_1$ is the number of edges, $N_2$ is the
number of squares (faces), and $N_3$ is the number of basic cubes in
the excursion set for a given filling factor (density threshold).
This formula was proven by \citet{adler} and was first used in
cosmological studies by \citet{coles96}; we refer to that paper for
a thorough discussion of the method and of necessary boundary
corrections.

This algorithm is simple to program, and it gives excellent results,
provided the grid step is substantially smaller than the
characteristic lengths of the isosurfaces (the smoothing length).
This is needed to be able to accurately follow the geometry of the
surface. It is also very fast, allowing the use of Monte-Carlo
simulations for error estimation.

\begin{figure}
\centering
\resizebox{0.5\textwidth}{!}{\includegraphics*{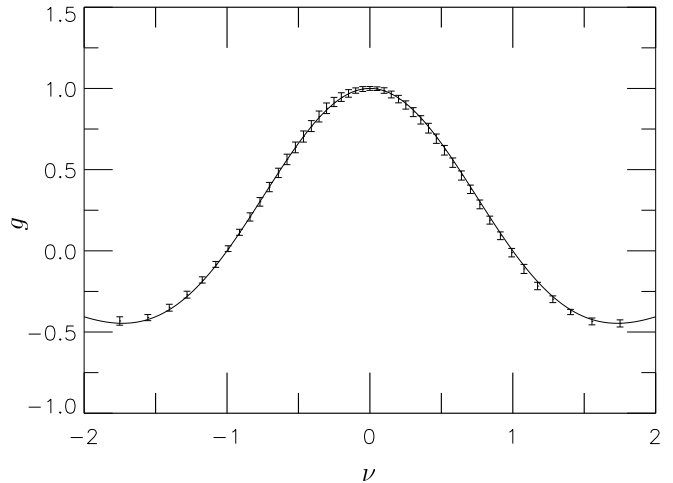}}
\caption{The average genus curve for 50 realizations of a Gaussian
random field with $P(k)\sim k^{-1}$ together with the expected
analytical result (solid line). The error bars show 1 $\sigma$
deviations. \label{grf50}}
\end{figure}

In order to test the algorithm and our program, we calculated the
genus curve for 50 realizations of a Gaussian random field  with a
power-law power spectrum $P(k)\sim k^{-1}$ in a $128^3$ box. The
realizations were smoothed with a Gaussian kernel of $\sigma=3$. The
results are shown in Fig.~\ref{grf50}. Our results are very close to
the theoretical expectations, and the errors are similar to those
reported recently by \citet{surfgen}, who used a very precise
algorithm based on triangulated surfaces (SURFGEN).

\section{Minkowski functionals of simulated point distributions}

In this section, we apply Gaussian smmothing and wavelet denoising procedures
to three different point sets. For the Gaussian kernel, we choose
different values of the bandwidth $\sigma$. The fourth Minkowski
functional (the Euler-Poincar\'e characteristic $V_3$) is then
calculated for the smoothed density fields. We will see how the
Gaussian smoothing tends to bring the $V_3$-curve closer to the
expected one for a Gaussian random field, independently of the
characteristics of the initial field. It demonstrates that the
morphological characteristics, obtained by Gaussian smoothing, may
carry more information about the filter itself than about the point
process. We have chosen different point processes with genuinely
non-Gaussian features, and with different topologies.

\subsection{Description of the samples}

The first data set used in this analysis has been generated by
A.~Klypin from an $N$-body simulation, and has been used in wavelet
applications before (see, e.g. Starck \& Murtagh 2002, p. 221). This
simulation is described by \citet{klypin97}. It contains 14616
galaxies within a cube of size of 60$h^{-1}$ Mpc. All the three
samples have similar number of data points, and we calculate the
Minkowski functionals for all three samples, using a $128^3$ mesh.
The correlation length $r_0$ and the size of the volume per
particle $d$ for this sample, both in physical and
grid units, are given in Table~\ref{tab:r0}. We also give the mean
nearest-neighbor distance for the sample ($\langle\mbox{nnd}\rangle$). 
If a sample is not too
heavily clustered, this should be close to $d$.

\begin{table}
\caption{Simulated point distributions.\label{tab:r0}}
\setlength{\tabcolsep}{5pt}
\begin{tabular}{lrrrrrrrr}\hline
&$N$&$L$&$r_0$&$d$&$r'_0$&$d'$&
    $\langle\mbox{nnd}'\rangle$\\
\hline
nbody&14616&60&4.0&2.45&8.5&5.2&4.5&\\
filaments&14718&100&10.0&4.1&12.8&5.2&2.6\\
cheese&14718&128&27.8&5.2&27.8&5.2&1.1\\
\hline
\end{tabular}

\small Note: The first three lengths ($L,r_0$ and $d$) are in units
of $h^{-1}$ Mpc, the last three lengths ($r'_0,d'$ and
$\langle\mbox{nnd}'\rangle$), in grid units.
\end{table}

The second point process is based on Voronoi tessellation. We
generate a Voronoi tessellation similar to the observed large-scale
galaxy distribution, with the mean size of cells of 40$h^{-1}$ Mpc
in a 100$h^{-1}$ Mpc cube, and populate the edges of the cells
(filaments). There are about 26 Voronoi cells; the sample. contains
14718 points, all close to filaments, with a $r^{-2}$ cross-section
density profile, and a 3$h^{-1}$ Mpc density scale. About 70--75\%
of the space is empty. Table~\ref{tab:r0} gives the characteristic
lengths for this sample.

We will call the third data set the ``Swiss cheese" model. In a
$128^3$ cube, we cut out 40 holes with radii $R$ in $[20,40]$, with
a uniform distribution of hole volumes. About 80--83\% of the
sample volume is empty, the remaining volume is filled with a
Poisson distribution of about 15000 points (see Table~\ref{tab:r0}).

\begin{figure}
\centering
\resizebox{0.5\textwidth}{!}{\includegraphics*{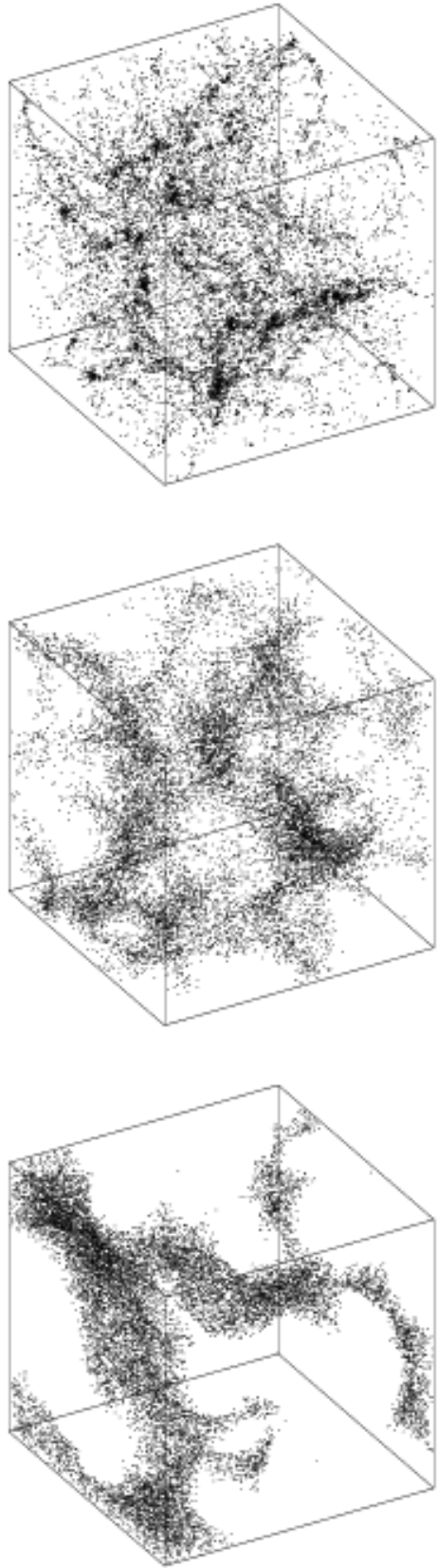}}
  \caption{The three data sets that will serve to illustrate the different smoothing
  schemes and  their implications when estimating the Euler characteristic. The top
  panel shows the $N$-body data, the middle panel shows the
  Voronoi filament model, and the bottom panel -- the nearly-empty Swiss cheese model.}
  \label{fig:points}
\end{figure}

These simulated galaxy distributions are shown in
Fig.~\ref{fig:points}.

\subsection{Smoothing and morphology}

\begin{figure*}
\centering \resizebox{\textwidth}{!}{\includegraphics*{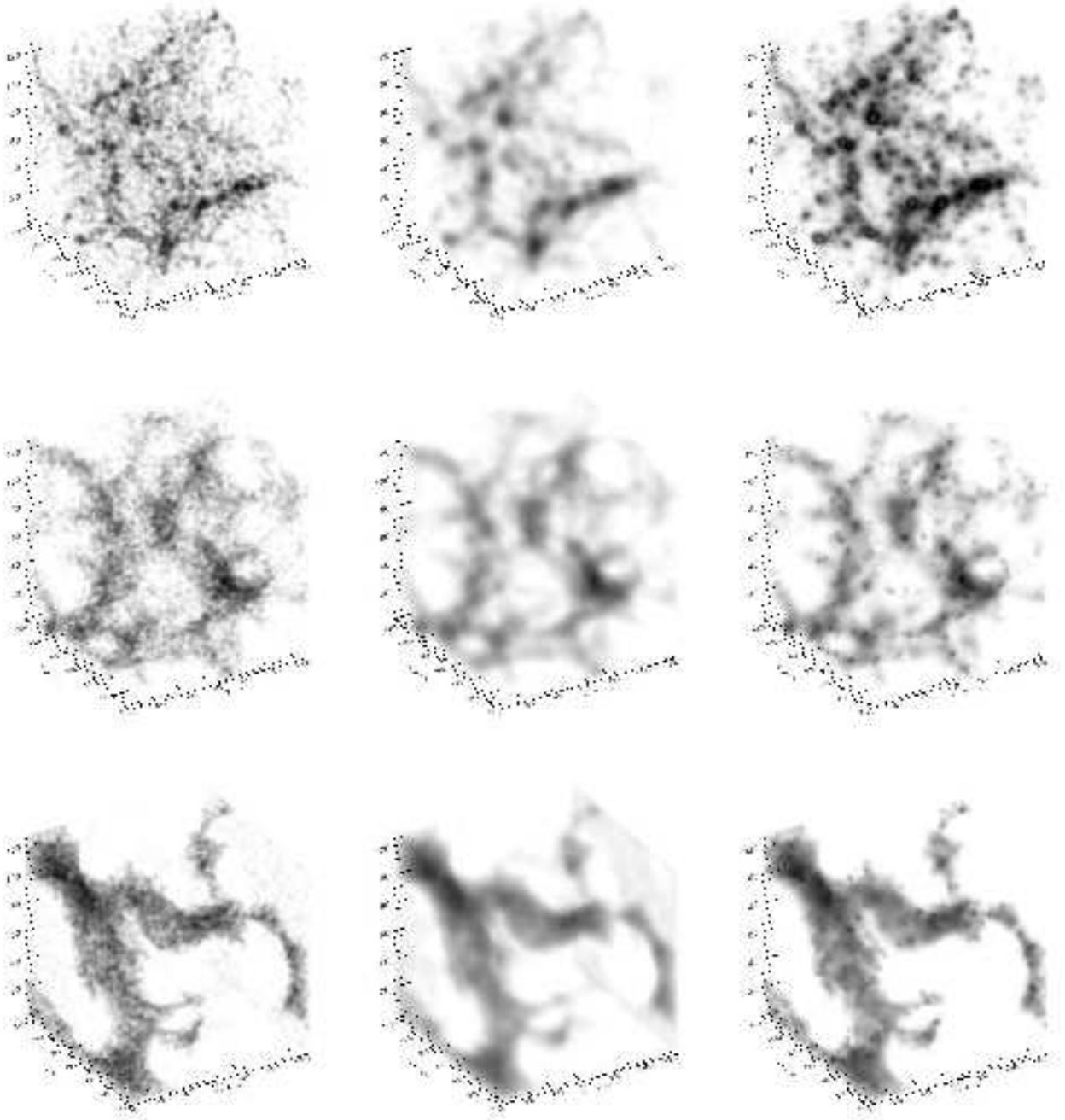}}
 \caption{Rendering of the density fields, obtained by smoothing of
 the three data sets shown in Fig. \ref{fig:points} with a Gaussian filter
 with $\sigma=1$ (first column), $\sigma=3$ (second column) by and wavelet
 denoising (third column). The smoothing lengths are given in grid units.}
    \label{fig:den}
\end{figure*}

In order to find the morphological descriptors (Minkowski
functionals) for our realizations of point processes we have to
smooth the data to obtain a continuous density field. The usual
approach is to use Gaussian kernels for smoothing; we shall compare
the results with those obtained by the wavelet-based smoothing
scheme introduced in this paper. We calculated all MF-s, but as the
functional $V_3$ shows more details than others, we show only the
results for this functional here. Fig.~\ref{fig:den} shows, in the
panels of the right column, the three point patterns of
Fig.~\ref{fig:points} filtered by the 3D wavelet transform, using
the algorithm described previously. The left and the middle panels
of each row correspond to Gaussian smoothing with $\sigma=1$ and
$\sigma=3$ (in grid units), respectively.

\begin{figure}
\centering
\resizebox{.48\textwidth}{!}{\includegraphics*{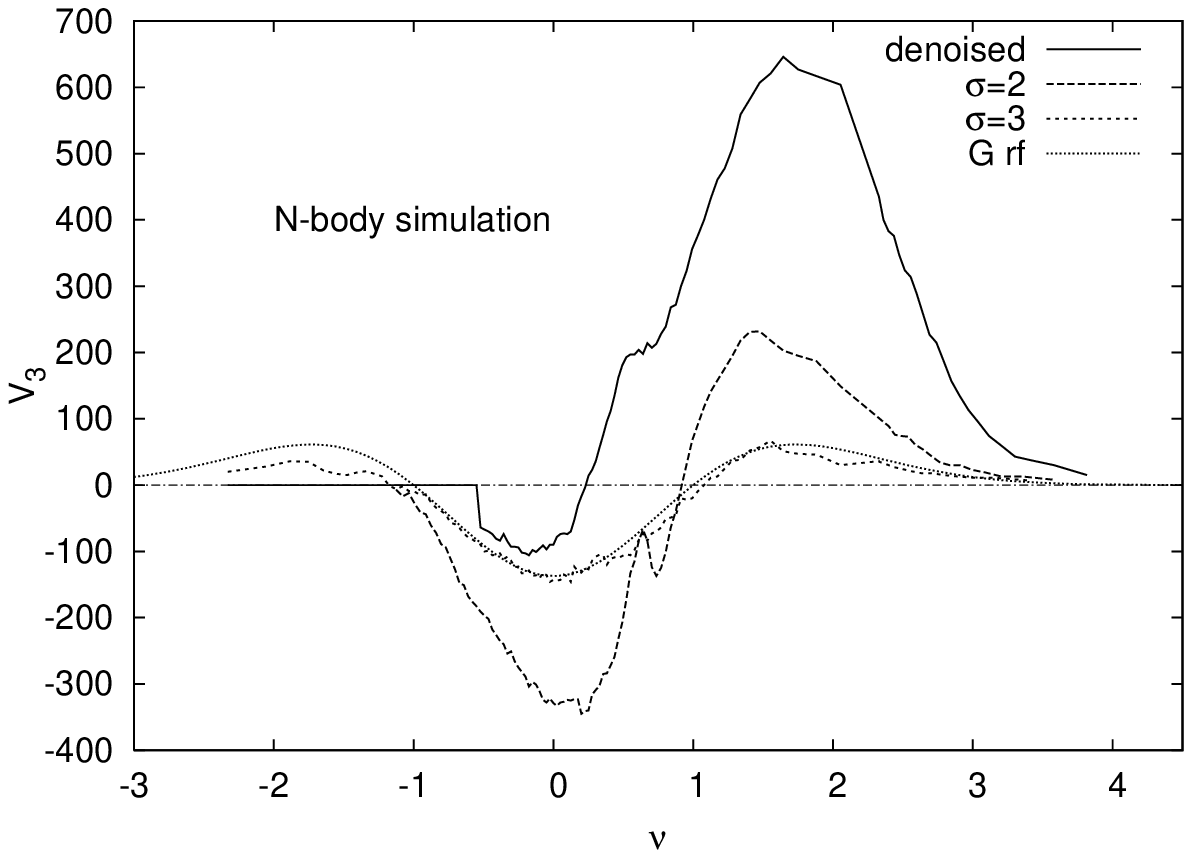}}
   \resizebox{.48\textwidth}{!}{\includegraphics*{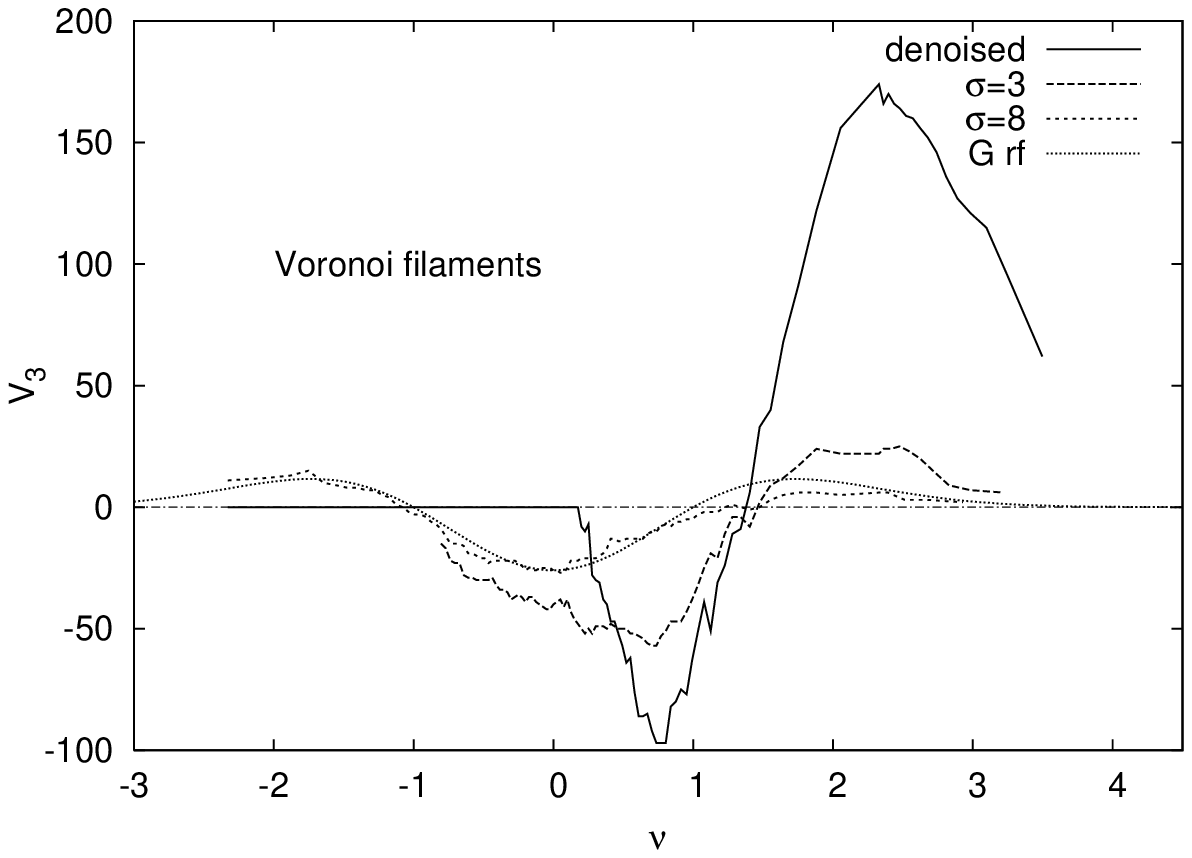}}
   \resizebox{.48\textwidth}{!}{\includegraphics*{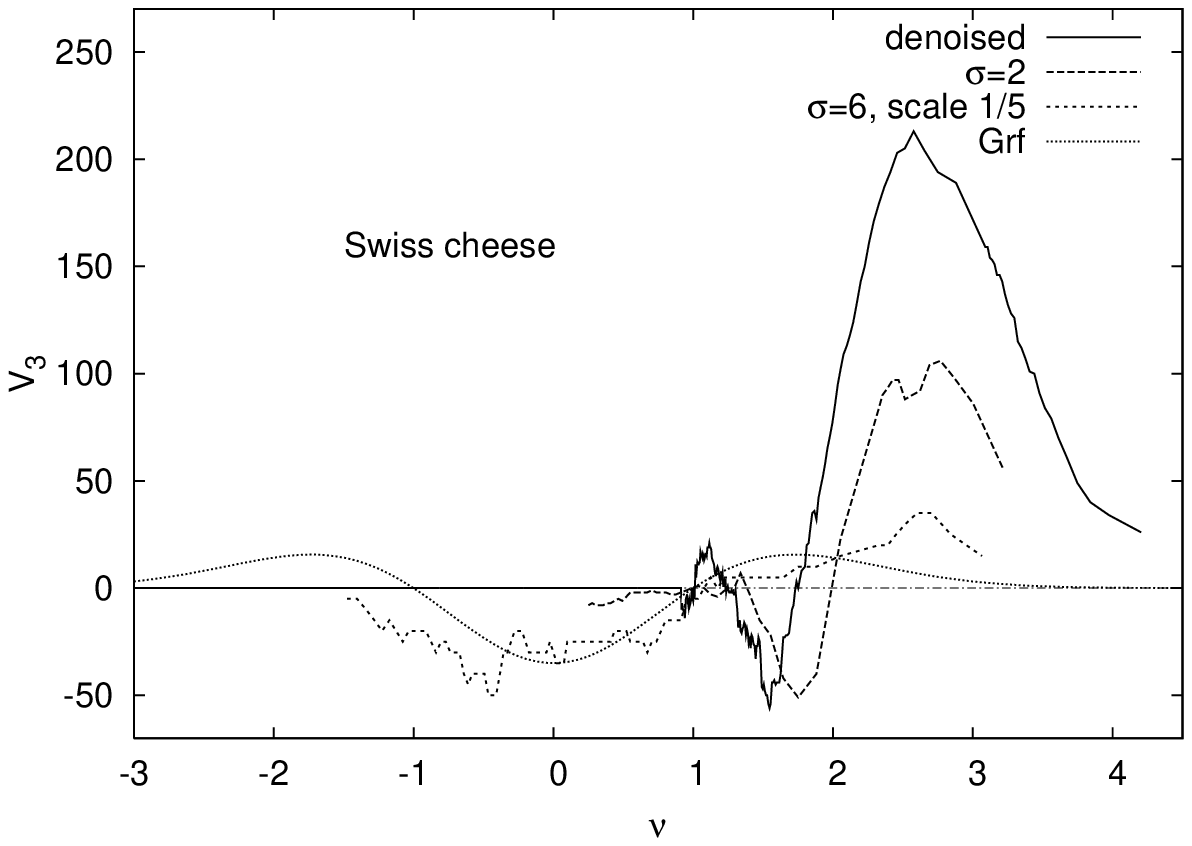}}
  \caption{The $V_3$ curves for the three point distributions. We show
in  each panel
  the curves obtained by smoothing the data with a Gaussian
  with two different filter widths (in grid units)
and the MF $V_3$ for the wavelet
filtered data set. As previously, the top
  panel corresponds to the $N$-Body simulation, the middle panel is for the
  Voronoi filament model, and the bottom panel corresponds to
  the nearly-empty Swiss cheese model.}
  \label{fig:genus}
\end{figure}

We can clearly see that when the bandwidth is too small (left
panels), discreteness and noise dominate the reconstructed density
fields, while using a larger value of $\sigma$ tends to erase all
the small scale features of the distribution. This is also shown in
Fig.~\ref{fig:genus}, where we can see that the morphology of the
Gaussian-smoothed density field, as described by the Euler
characteristic $V_3$, depends strongly on the width of the Gaussian
filter. This width is a free parameter and thus the
Gaussian-filtered density field is not uniquely determined. Choosing
the width of the filter we discard information on scales of that
width and smaller. On the other hand, the wavelet transform 
leads to a sparse representation of the density field
and allows us to detect and keep at all scales  coefficients   
which have the greatest probability to be real. 
This is demonstrated by the 3D image in the right panels of
Fig.~\ref{fig:den}, where we see, e.g., in the rendering of the
$N$-body model (top-right) how large filaments, big clusters and
walls coexist with small scale features such as the density
enhancements around groups and small clusters. The Euler
characteristic of this adaptive reconstructed density field is much
more informative, because it is unique, it does not depend on the
particular choice of the filter radius. Because of that, wavelet
morphology is clearly a more useful tool than the usual approach of
Gaussian smoothing. Also, the Minkowski functionals of
Gaussian-smoothed density fields mimic those of Gaussian random
fields, in contrast with the wavelet-based approach. Thus, they
describe more the properties of the filter, than the real morphology
of the density distribution.

This is seen already in the case of the $N$-body model (the top
panel of Fig.~\ref{fig:genus}), where the $V_3$ curve is close to
Gaussian already for $\sigma=3$ (in grid units), much smaller than
$r_0$, and even smaller than the mean nearest-neighbor distance.

\begin{figure*}
\centering
\resizebox{\textwidth}{!}{\includegraphics*{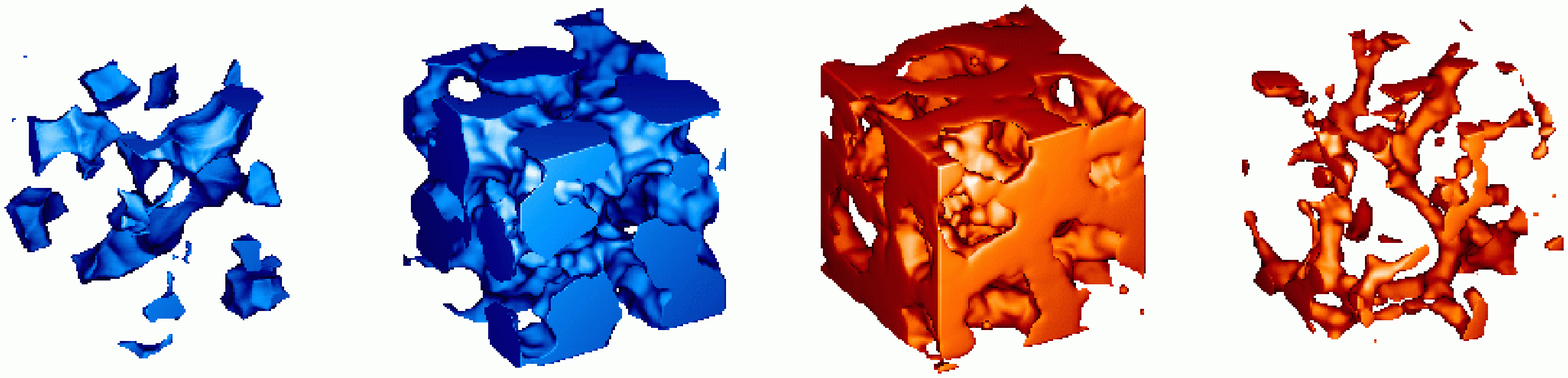}}
   \resizebox{\textwidth}{!}{\includegraphics*{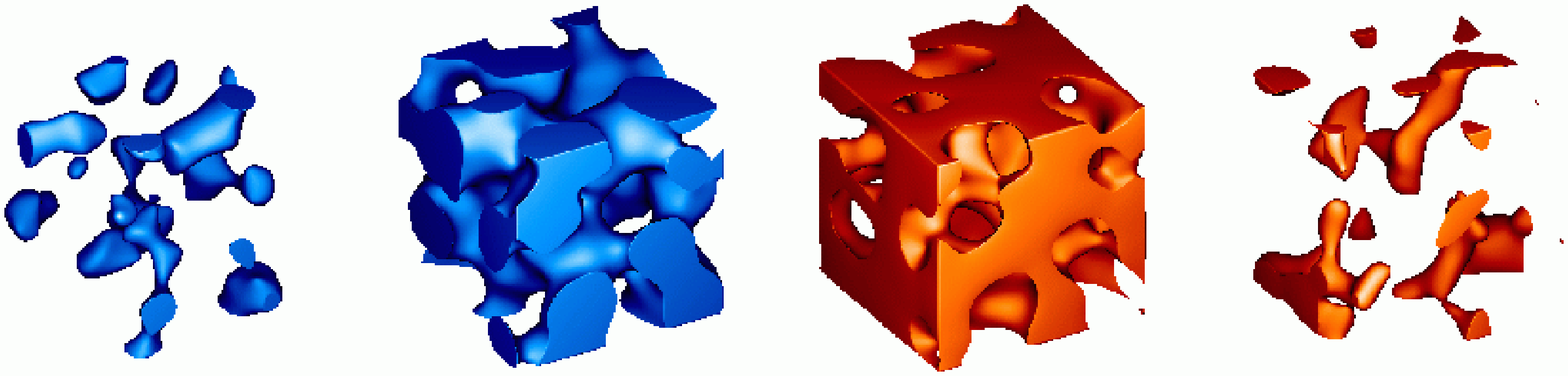}}\\
   \resizebox{\textwidth}{!}{\includegraphics*{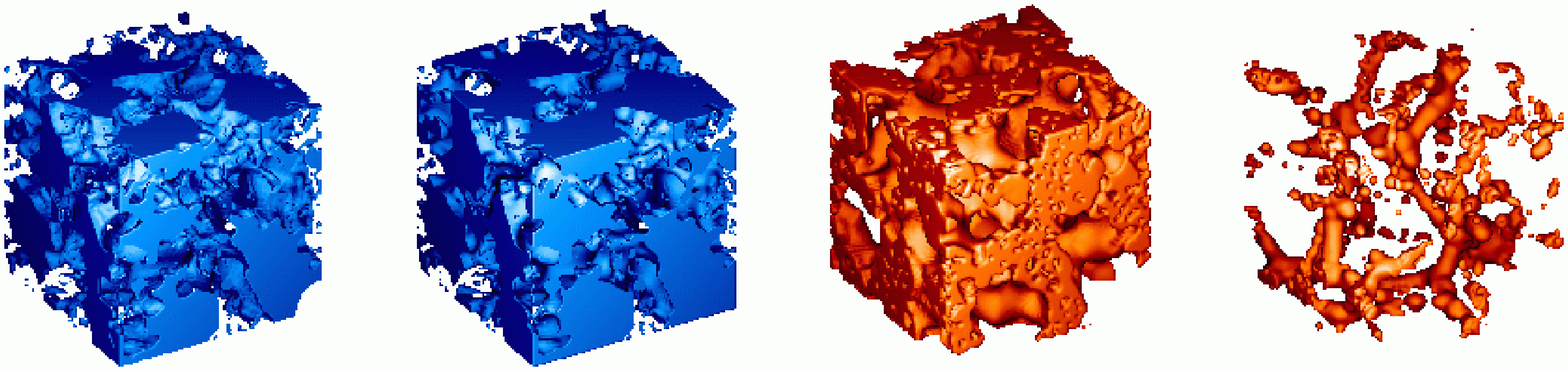}}\\
\caption{The isodensity surfaces corresponding to the Voronoi
filament model for the Gaussian-smoothed field with $\sigma=3$
(upper row), $\sigma=8$ (middle row, all in grid units) and the
wavelet-denoised field (bottom row). The density thresholds
delineate, from left to right, 7\% low, 50\% low, 50\% high and 7\%
high density regions.} \label{fig:filiso}
\end{figure*}

For the clearly non-Gaussian Voronoi filament model, when we
increase the value of $\sigma$, the $V_3$ curves also approach the
typical shape for a Gaussian field (see the middle panel of
Fig.~\ref{fig:genus}), while the wavelet-denoised  density shows the
expected behavior for the Euler characteristic for this kind of
spatial configuration. The three curves shown in the middle panel of
Fig.~\ref{fig:genus} correspond to the iso-density contours shown in
Fig.~\ref{fig:filiso}. While for Gaussian smoothing with $\sigma=3$
it is still possible to see the filamentary structure in the $V_3$
diagram, for $\sigma=8$ the isocountours are indistinguishable of
those of a Gaussian field like the one shown in
Fig.~\ref{fig:gaussden}. It is clear that such a smoothing is
excessive, and destroys the original morphology of the point sample.
The $V_3$ curve is, in fact, close to Gaussian for $\sigma=6$
already. Both $\sigma=6$ and $\sigma=8$ are smaller than the
correlation length of this sample (see Table~\ref{tab:r0}), and
$\sigma=6$ is close to the size of the volume-per-particle $d$.

The nearly-empty Swiss cheese model is even more non-Gaussian, and
therefore, even for large values of $\sigma$, Gaussian smoothing
does not converge to the symmetric $V_3$ curve. Nevertheless, the
shape of the curve for the Gaussian-smoothed density depends
strongly on the bandwidth, and again the curve for the
wavelet-denoised density is clearly more representative of the true
underlying morphology.

\section{Morphology of the 2dFGRS}

\subsection{Data}

The best available redshift catalog to study morphology of the
galaxy distribution at present is the 2dF Galaxy Redshift Survey
(2dFGRS) \citep{2df}. It fills large compact volume(s) in space and
includes more than a quarter of million of galaxies.  This is a
flux-limited catalog and therefore the density of galaxies decreases
with distance. For statistical analysis of such of surveys, a
weighting scheme that compensates for the missing galaxies at large
distances, has to be used. Usually, each galaxy is weighted by the
inverse of the selection function \citep{cf:martinezsaar}. However,
the resulting densities will have different resolution at different
locations, and will not be suitable for morphological studies.

At the cost of discarding many surveyed galaxies, one can
alternatively use volume-limited samples. In this case, the
variation in density at different locations depends only on the
fluctuations of the galaxy distribution itself. We have used the
volume-limited samples prepared by the 2dF team for scaling studies
\citep{croton1,croton2}, and kindly sent to us by Darren Croton. As
our basic sample, we chose the catalog with absolute luminosities in
the range $-19 > M_{B_J} -5\log_{10} h > -20$ (the type dependent
$k+e$ correction \citep{norberg02} has been applied to the
magnitudes). This sample contains galaxies with luminosity around
$L_\ast$. This catalog is the largest of the 2dF volume-limited
catalogs, and as \citet{baugh04}  point out, it provides optimal
balance between the surveyed volume and the number density of
galaxies. Although the catalog does not suffer from luminosity
incompleteness, it is slightly spectroscopically incomplete, mainly
due to missing galaxies because of fiber collisions. The
incompleteness parameter has been determined by every galaxy by the
2dF team; when calculating densities, each galaxy can be weighted by
the inverse of this parameter.

We split the volume-limited sample into the Northern and Southern
subsamples, and cut off the numerous whiskers in the plane of the
sky to obtain compact volumes.

We performed morphological analysis for both the Southern and
Northern subsamples. The grid-based scheme we use works well for
simple cuboid geometries. The geometry of the Northern sample is
similar to a flat slice, while the Southern sample is enclosed
between two cones of opening angles of $64.5^\circ$ and
$55.5^\circ$.  When we tried to cut cuboidal volumes (bricks) from
the Southern sample cone, we ended up with small brick volumes.  So
we carried out the morphological analysis for the full volume of the
Southern sample, only to find that the border corrections for the
Minkowski functionals are large and uncertain. Thus we report in
this paper only the results of the analysis for the Northern sample.

In order to obtain a compact volume, we choose the angular limits
for the Northern sample as $-4.5^\circ \le \delta \le 2.5^\circ$ and
$ 149.0^\circ \le \alpha \le 209.0^\circ$. The slice lies between
two cones defined by the $\delta$ limits. The right ascension limits
cut the cones by planes from both sides, and there are two
additional cuts by two spheres. The radii of the spheres are fixed
by the original data, and depend only on the chosen absolute
magnitude limits (and on the cosmological model). For our sample
they are: $R_1=61.1$ $h^{-1}$ Mpc, $R_2=375.6$ $h^{-1}$ Mpc.

\begin{figure}
\centering
\resizebox{0.5\textwidth}{!}{\includegraphics*{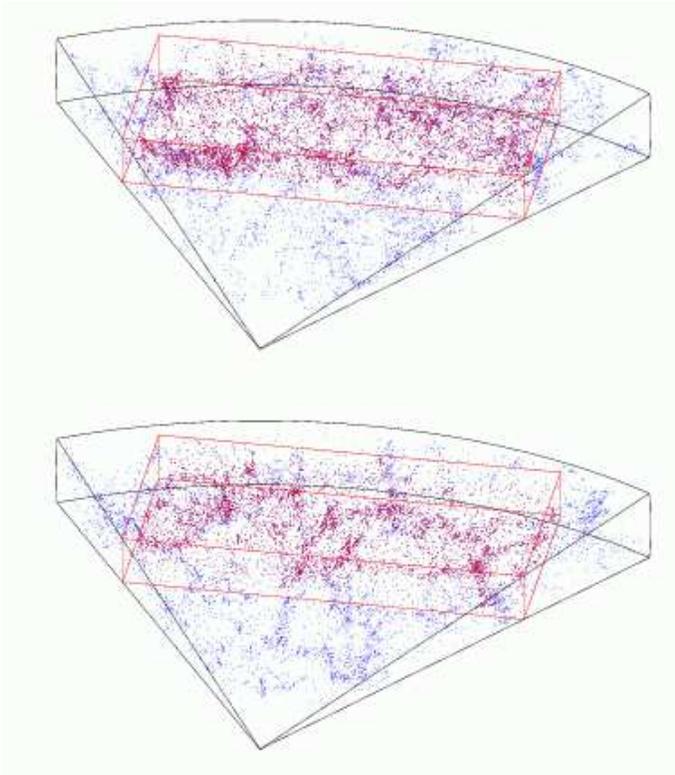}}
  \caption{The volume-limited cuboidal sample analyzed in this paper
  drawn from the Northern slice of the 2dFGRS (top) and from
  a mock realization.}
  \label{fig:2dfpoints}
\end{figure}

    As this sample is pretty flat, we cut from it a maximal volume
cuboidal window, a ``brick" with dimensions of $254.0 \times  133.1
\times 31.1$ $h^{-1}$ Mpc, with 8487 galaxies (see
Fig.~\ref{fig:2dfpoints}). This gives for the per-particle-volume
size $d=5.0\,h^{-1}$ Mpc.

\subsection{Mock catalogs}

In order to estimate sample errors of the Minkowski functionals, we
use mock catalogs, provided by the 2dF team. \citet{norberg02}
created 22 mock catalogs for the 2dFGRS that have been used by the
2dfGRS team to measure the influence of cosmic variance of different
statistics, as correlation functions, counts-in-cells, the void
probability function, clustering of groups, etc.
\citep{croton1,croton2,baugh04, padilla04}. The mock catalogs were
extracted from the Virgo Consortium $\Lambda$CDM Hubble volume
simulation, and a biasing scheme described in \citet{cole98} was
used to populate the dark matter distribution with galaxies. The
catalogs were created by placing observers in the Hubble volume,
applying the radial and angular selection functions of the 2dFGRS,
and translating the positions and velocities of galaxies into
redshift space. No luminosity clustering dependence is present in
the mock catalogs.

\begin{figure}
\centering
\resizebox{0.5\textwidth}{!}{\includegraphics*{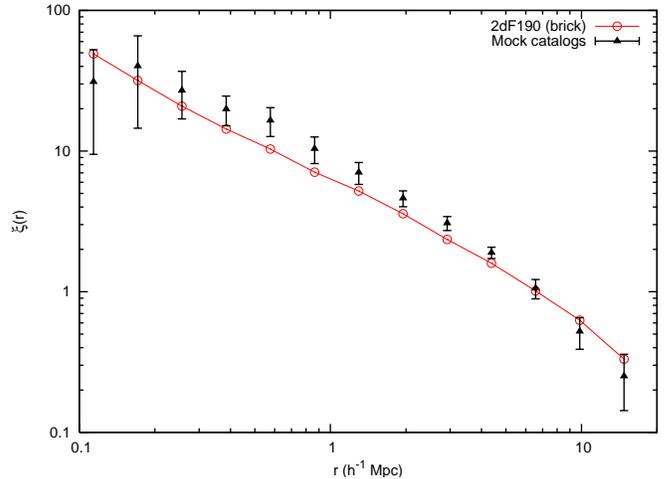}}\\
  \caption{The two-point correlation function of the 2dF brick
  (open circles) together with the average and the total
deviation range for the 22 mock catalogs.}
  \label{fig:xi}
\end{figure}

The mock catalogs represent typical volumes of space. The real 2dF
catalog, however, includes two superclusters, one in the Northern,
another in the Southern subsample (see a thorough discussion in
\citet{croton2}). The Northern supercluster is especially prominent
in the $M\in[-19,-20]$ survey; all mock samples for this catalog
have less galaxies than the 2dF sample. We cut mock bricks from the
mock samples, too, as we did for the real 2dF data; the mean number
of galaxies in the mock bricks is 1.36 times smaller than in the 2dF
brick. The supercluster shows up in the correlation function, too,
enhancing correlations at intermediate scales, compared to those of
the mocks (Fig.~\ref{fig:xi}). The correlation length for the brick
is $r_0=6.8\,h^{-1}$ Mpc, only slightly larger than the
characteristic length $d=5.0\,h^{-1}$ Mpc. We remind the reader that this is
the correlation length for redshift space; the 2dF correlation
length for real space has been estimated as $R_0=5.05\,h^{-1}$ Mpc
\citep{hawkins03}. The mean nearest-neighbor distance is 2.3
$\,h^{-1}$ Mpc, showing that the galaxy distribution is well clustered.

\subsection{Minkowski functionals of the 2dFGRS Northern sample}

As we said, we show the results for only one volume-limited
subsample of the 2dFGRS Northern area. Other subsamples have either
smaller volumes or smaller galaxy densities.

We do not use the weights to correct for spectroscopic
incompleteness for the final results. 
We have seen that the influence of the weights in the
correlation function $\xi(r)$ is negligible. A similar test
has been performed by \citet{croton1} using counts-in-cells
statistics on mock catalogs, both complete and incomplete, reaching
a similar conclusion.
We have tested the influence of the incompleteness
by calculating density fields for several Gaussian smoothing
lengths with and without the weights, and compared the
resulting Minkowski functionals. The differences were almost 
imperceptible, thus we decided for the conceptually simpler 
procedure.

We calculate the Minkowski functionals by sweeping over the grid (we
use a 1$h^{-1}$ Mpc grid step). We start at the nearby border
planes, and we account for the edge effects for bricks by not using
the grid vertices at the faraway borders. We tested this procedure
by using realizations of Gaussian random fields; although the border
effects are small, the correction works well. We estimate the
significance of the deviations of the MF curves from those for a
Gaussian random field, by calculating them for a large number of
Gaussian realizations (about 1300). In order to create these
realizations, we adopted the analytical approximation for the power
spectrum by \citet{klypin97}, for parameters similar to the
concordance model ($\Omega_{\mbox{\scriptsize matter}}=0.3,
\Omega_{\Lambda}=0.7,\Omega_{\mbox{\scriptsize bar}}=0.026, h=0.7$).

In order to estimate the cosmic variance, we use the 22 mock bricks
described above. As the distribution of MF amplitudes is rather
asymmetric, we do not find the variance, but we show the total range
of variation of the mock MF curves. As there are 22 mock samples,
this range is close to the usual Gaussian ``2 sigma'' confidence
regions. Thus, the confidence regions for Gaussian realizations
given in the figures, are also given for the $2\sigma$ (95\%) level.

We noticed above that the mock catalogs miss the supercluster present in the
real 2dF sample (look at the front left region of the 2dF brick in
the upper panel of Fig.~\ref{fig:2dfpoints}), and have
systematically lower density than the real 2dF sample. The fix
adopted by \citet{croton2} was to cut out the region surrounding the
supercluster. We cannot do that, as this would lead to complex
boundary corrections. For wavelet cleaning this should not be a
problem, the algorithm will automatically follow the density
distribution. For Gaussian smoothing, we compensated for the density
difference by using 1.11 times wider smoothing lengths for mocks
than for the 2dF brick. The smoothing lengths for the Gaussian
realization remain unscaled, of course.

\begin{figure}
\centering
\resizebox{.48\textwidth}{!}{\includegraphics*{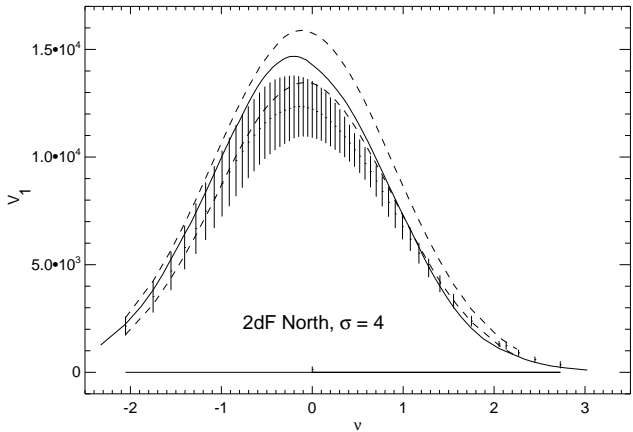}}
\caption{The Minkowski functional $V_1$ for the 2dF GRS Northern brick, 
for Gaussian smoothing with $\sigma=4h^{-1}$Mpc (solid line). The cosmic
error is characterized by the variability of $V_1$ for 22 mock samples
(shown by bars, the same smoothing). The 95\% confidence regions 
for the theoretical prediction, $\sigma=4h^{-1}$Mpc-smoothed 
realizations of Gaussian random fields with the 'concordance cosmology' 
power spectrum, are shown by dashed lines.
\label{fig:2dfv1}}
\end{figure}

We start with the first two nontrivial Minkowski functionals (the
first MF, $V_0$, is trivially Gaussian due to our choice of the
argument $\nu$). The second (Fig.~\ref{fig:2dfv1}) MF (the area of
the isodensity surfaces) for the Gaussian smoothing with $\sigma=4$
(grid units or $h^{-1}$ Mpc) barely fits into the 95\% Gaussian
confidence interval (it lies completely in the $3\sigma$ interval).
It is interesting that the values of $V_1$ for the mocks lie mostly
outside of it -- the isodensity surfaces are smoother than for the
real data (recall the supercluster), and than for the Gaussian
random field, too.

\begin{figure}
\centering
\resizebox{.48\textwidth}{!}{\includegraphics*{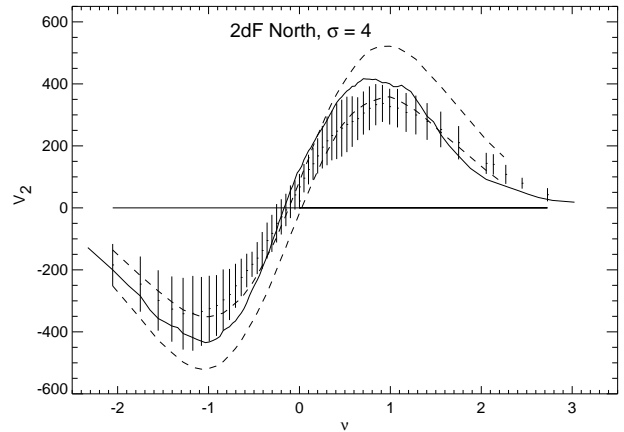}}
\caption{The Minkowski functional $V_2$ for the 2dF GRS Northern brick, 
for Gaussian smoothing with $\sigma=4h^{-1}$Mpc (solid line). The cosmic
error is characterized by the variability of $V_2$ for 22 mock samples
(shown by bars, the same smoothing). The 95\% confidence regions 
for the theoretical prediction, $\sigma=4h^{-1}$Mpc-smoothed 
realizations of Gaussian random fields with the 'concordance cosmology' 
power spectrum, are shown by dashed lines.
\label{fig:2dfv2}}
\end{figure}

The third (Fig.~\ref{fig:2dfv2}) MF (the mean curvature of the
isodensity surfaces) for the Gaussian smoothing with the same
$\sigma=4$ as above also lies a bit outside of the 95\% Gaussian
confidence interval, but fits completely in the $3\sigma$ interval,
not shown in the figure. Mocks do not lie well within the 95\% 
confident Gaussian
band, while the $V_2$ curve for the 2dF data lies
close to the extreme $V_2$ values of the
mock catalogues shown by bars in the diagram.  
These two figures show that Gaussian smoothing with
$\sigma=4$ (recall that $r_0=6.8 \,h^{-1}$ Mpc for the 2dF brick)
has already given a nearly Gaussian morphology to the data.

\begin{figure}
\centering
\resizebox{.48\textwidth}{!}{\includegraphics*{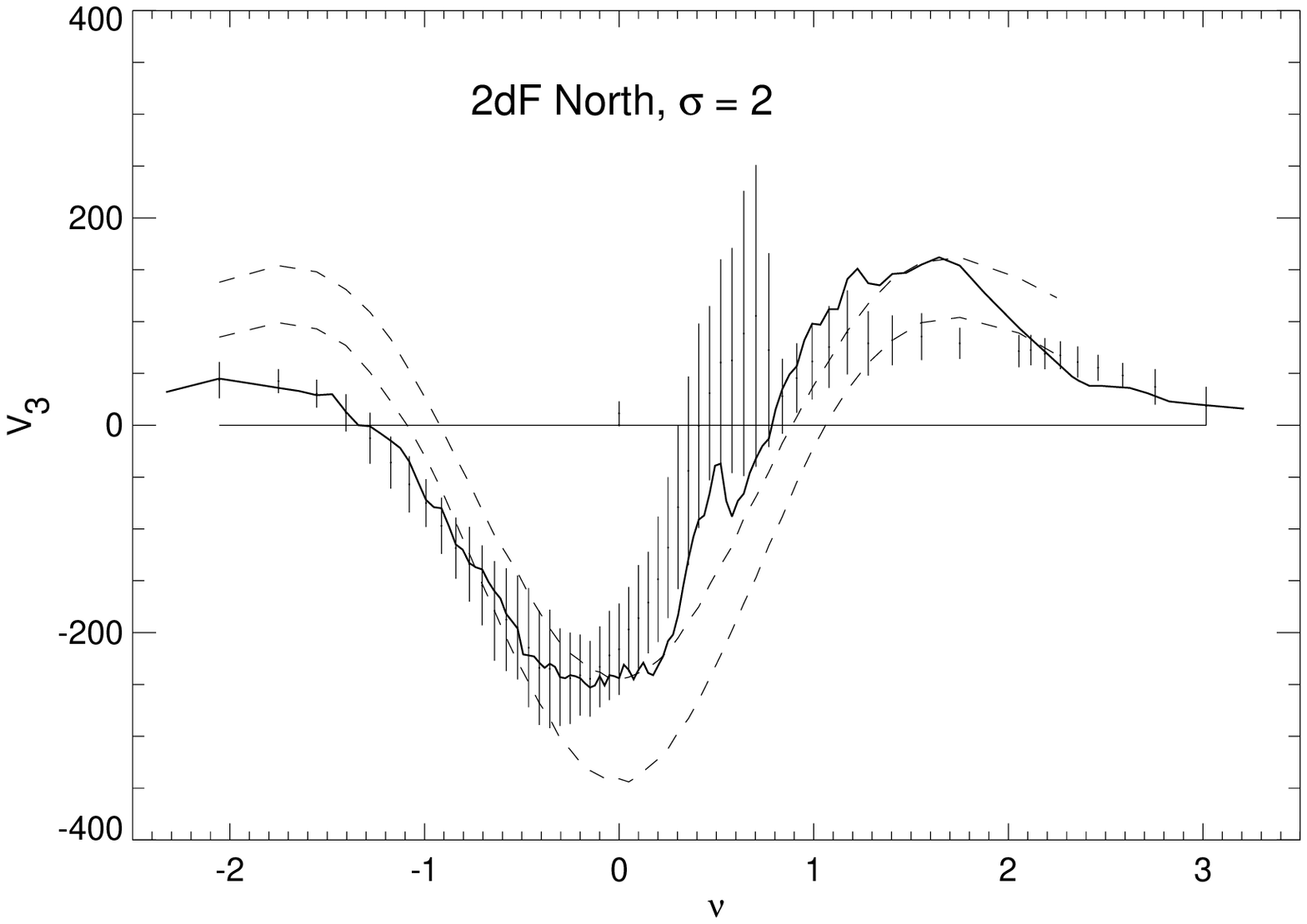}}
   \resizebox{.48\textwidth}{!}{\includegraphics*{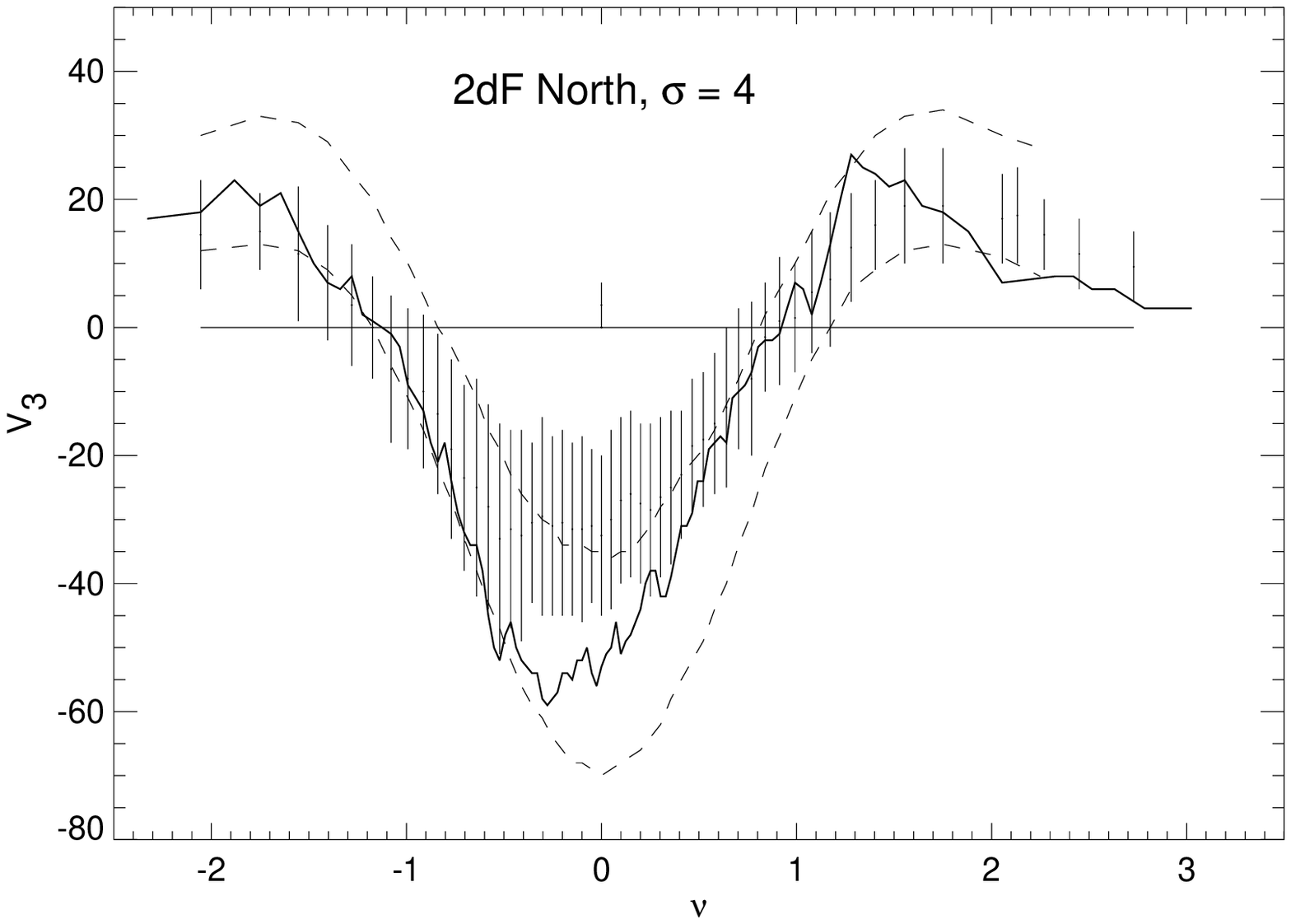}}
   \resizebox{.48\textwidth}{!}{\includegraphics*{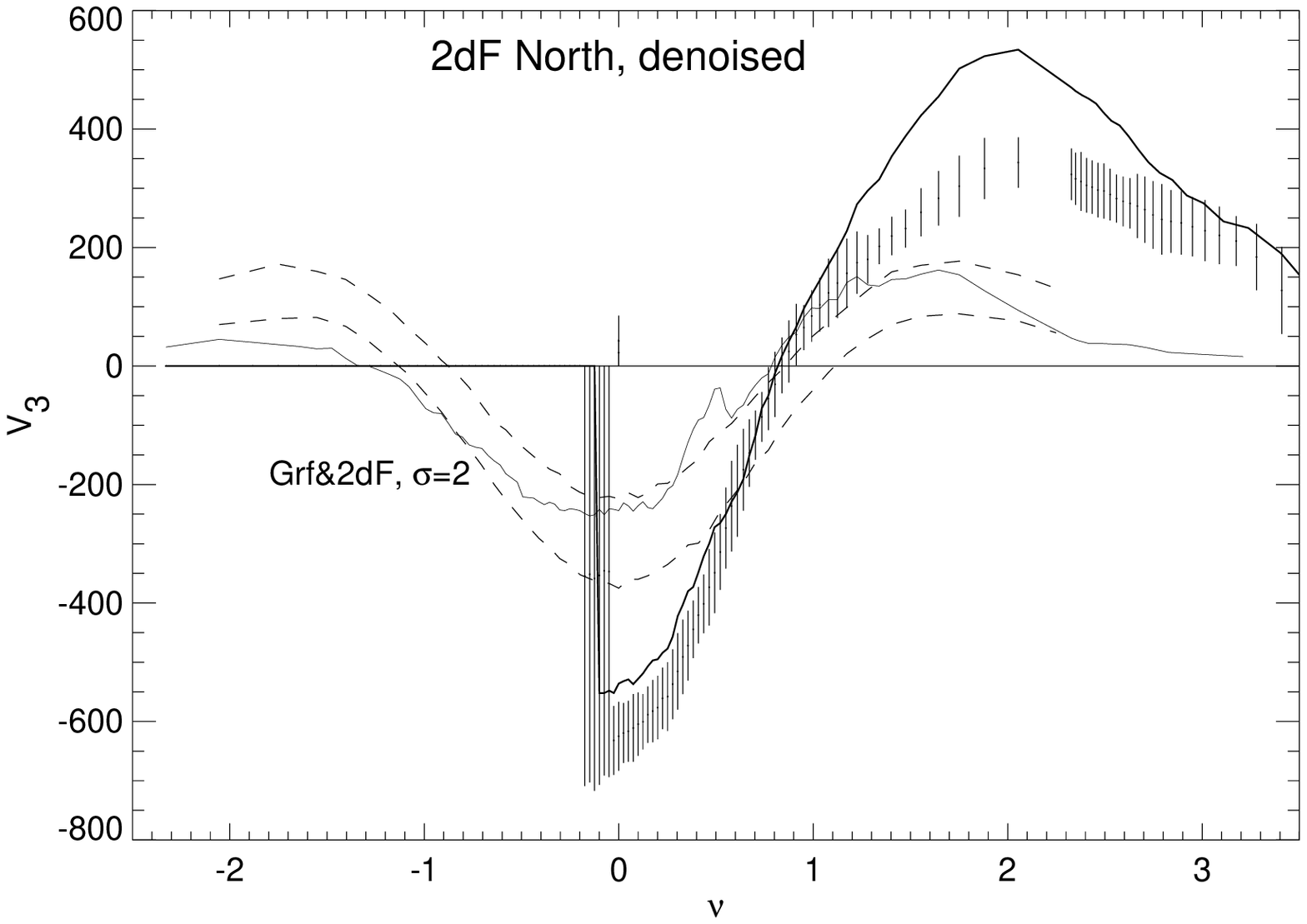}}
  \caption{The Minkowski functional $V_3$ for the 2dF brick. The upper
panels show the results for Gaussian smoothing with $\sigma=2h^{-1}$Mpc
and $\sigma=4h^{-1}$Mpc, respectively (the designations are the same as
in the previous two figures). The bottom panel describes wavelet morphology
of the 2dF GRS, showing the $V_3$ curve for the wavelet denoised data
set (thick solid line), and comparing it with the variability range of the
wavelet denoised mocks (bars). We show also the 95\% confidence limits
for 1300 realizations of theoretical Gaussian density fields (dashed lines),
and the $V_3$ data curve (thin solid line), all obtained for the 
Gaussian $\sigma=2h^{-1}$Mpc smoothing.
\label{fig:2dfv3}}
\end{figure}

As usual, the $V_3$ curves (Fig.~\ref{fig:2dfv3}) show the most
details. The upper panel shows that the data smoothed with a
Gaussian filter of width $\sigma=2$, is yet undersmoothed, but does
not differ very much from a Gaussian random field. Discreteness
effects are more evident for the mock samples (the peak around
$\nu=0.7$). The middle panel demonstrates again that the density
field smoothed with a Gaussian filter of width $\sigma=4$ can
already be considered Gaussian, and the mocks do not differ much
from Gaussian realizations either.

These two panels show how the answer to the question of whether the density
distribution has intrinsically Gaussian morphology, depends on the
adopted smoothing widths. The bottom panel shows the result for
wavelet filtering of the point distribution. This curve is clearly
non-Gaussian, showing the presence of compact clusters for high-density
isosurfaces, and a spongelike morphology near $\nu=0$. However, in
contrast to the Gaussian case, the curve returns to 0 for smaller
values of $\nu$ -- about half of the sample space remains empty
after wavelet denoising. Gaussian smoothing, on the contrary, tends to
fill up the space.  The wavelet-filtered mocks show, in principle,
similar behavior to the data. They are only smoother, as seen from
the differences around $\nu=2$. It is interesting that the
wavelet-filtered $V_3$ curve is similar to those for the Voronoi
filament sample -- both samples are filamentary at larger scales.
Wavelet morphology returns a clear picture of the density field,
again, in contrast to the Gaussian-smoothed $V_3$ for the 2dF data,
where filamentarity is difficult to see.

\section{Conclusions}

We have presented a new wavelet-based method to study the morphology
of the galaxy distribution -- wavelet morphology. As we have shown,
it gives a unique morphological description, and is more accurate,
capturing the details of the distribution that are destroyed by
usual Gaussian smoothing. The code for the analysis of
wavelet morphology will be made available at
{\tt http://jstarck.free.fr}.

Using special highly non-Gaussian realizations of point processes,
we have demonstrated that Gaussian smoothing introduces Gaussian
features in the morphology, and is thus not the best tool to search
for departures from  Gaussianity.

We performed wavelet-morphological analysis of the most detailed 2dF
GRS volume-limited sample and found that it is clearly non-Gaussian.
The wavelet Minkowski functional $V_3$ finds high-density clusters,
large-scale filamentarity, and huge empty voids. A similar
morphological analysis, based on Gaussian smoothing, leads to the
conclusion that the morphology of the sample is close to Gaussian,
already for comparatively small smoothing lengths
($\sigma\ge4\,h^{-1}$ Mpc). This is a clear example of Gaussian
contamination.

The isotropic wavelet transform is optimal only for the detection 
of isotropic features, but
not for the detection of filaments or walls. A clear 
improvement could be made by using 
simultaneously several other multiscale transforms 
such the ridgelet transform and 
the beamlet transform which are respectively  well 
suited for walls and filaments \citep{sta05_3}.
This will be done in the future.

Wavelet morphology detects also the large supercluster in the
2dFGRS Northern sample, that has not been modeled by $N$-body mock
catalogs. A signature of the presence of this supercluster could
be  deduced from the correlation function. 
Gaussian morphology does not detect this feature.

\acknowledgements
We thank Rien van de Weigaert for his Voronoi program, 
Darren Croton for providing us with the 2dF volume-limited data
and explanations and suggestions on the manuscript, 
and Peter Coles for discussions.
This work has been supported by the University of Valencia
through a visiting professorship for Enn Saar,
by the Spanish MCyT project
AYA2003-08739-C02-01 (including FEDER), by the Generalitat
Valenciana project GRUPOS03/170, by the National Science 
Foundation  grant DMS-01-40587 (FRG),
and by the Estonian Science Foundation grant 6104.

\end{document}